\documentclass[%
 aip,
 amsmath,amssymb,
 reprint,%
]{revtex4-1}
\usepackage{xcolor}
\usepackage{graphicx}% Include figure files
\usepackage{dcolumn}% Align table columns on decimal point
\usepackage{bm}% bold math
\usepackage{setspace}
\usepackage[utf8]{inputenc}
\usepackage[T1]{fontenc}
\usepackage{mathptmx}
\usepackage{etoolbox}
\usepackage{amsmath} 
\usepackage{hyperref}
\usepackage{array}
\usepackage{amsmath, amssymb, xcolor}
\usepackage{mathtools}
\makeatletter
\def\@email#1#2{%
 \endgroup
 \patchcmd{\titleblock@produce}
  {\frontmatter@RRAPformat}
  {\frontmatter@RRAPformat{\produce@RRAP{*#1\href{mailto:#2}{#2}}}\frontmatter@RRAPformat}
  {}{}
}%
\makeatother

\begin{document}

\preprint{AIP/123-QED}

\title[ ]{Measuring Cyclic Tensile Properties of Fluids with Composite Harmonic Exponential Waveforms (CHEW)}
% Force line breaks with \\
\author{L.A. Kroo}
\affiliation{Massachusetts Institute of Technology, Department of Mechanical Engineering}
 \altaffiliation[L.A.K. Current affiliation: ]{Polymer Science and Engineering Department, University of Massachusetts Amherst}%Lines break automatically or can be forced with \\
\author{R.A. Nicholson}%
\affiliation{ 
Motif FoodWorks%\\This line break forced with \textbackslash\textbackslash
}%
\author{M.W Boehm}%
\affiliation{ 
Motif FoodWorks%\\This line break forced with \textbackslash\textbackslash
}%
\author{S.K. Baier}%
\affiliation{ 
Motif FoodWorks%\\This line break forced with \textbackslash\textbackslash
}%
\author{P.A. Underhill}
\affiliation{Rensselaer Polytechnic Institute, Department of Chemical Engineering}
 \email{lkroo@umass.edu, gareth@mit.edu}

\author{G.H. McKinley}
\affiliation{Massachusetts Institute of Technology, Department of Mechanical Engineering}

\date{\today}% It is always \today, today,
\begin{abstract}
Building off recent advances on how to practically use exponential shear in a torsional rheometer to compute transient planar extensional viscosity (\citet{kroo2025unifying}), we extend the technique to cyclic tensile measurements in complex fluids and soft solids. An input strain waveform of the form $\gamma(t) = 2 \sinh(\hat{\alpha} \sin(\omega t))$ provides a unifying approach that smoothly interpolates between exponential shear (ES) and oscillatory shear (SAOS/MAOS/LAOS) as the parameter $\hat{\alpha}$ is varied. Analogous to cyclic tensile fatigue tests in solids, or the process of chewing in the oral cavity, this complex strain history is used to quantify the evolution of extensional material properties at large strains over sequential cycles of stretch. In the limit of large values of $\hat{\alpha}$, the strain waveform locally increases exponentially and generates a period of strong material stretching with a Hencky strain rate given by $\hat{\alpha} \omega$. This allows for the direct computation of a transient planar extensional viscosity within specific domains of the periodic function. We demonstrate this technique on a set of model fluids, and then apply it to complex multiphase materials that “mutate”. These latter fluids exhibit progressive evolution in their rheological properties over repeated cycles of extensional deformation. Here we focus on two examples: a delicate foodstuff material (melted provolone cheese) which systematically decreases its extensional response over successive stretching cycles, mutating at a rate that is directly dependent on the effective Hencky strain rate. We contrast this with a PVA-borax solution which exhibits precisely the opposite effect during successive stretching cycles: increasing its planar extensional response over successive cycles, as interchain associative interactions (controlled via stretching) build structure within the fluid. These results highlight a promising new approach to study bulk extensional properties during cyclical stretching of complex fluids.

\end{abstract}

\maketitle

\begin{quotation}

This article summarizes a new technique for cyclically performing stretching flows on complex fluids in a torsional strain-controlled rheometer, by using a novel transient waveform. This waveform combines oscillatory flow with exponential shear and can be tuned smoothly between these two flow-type limits. 

%In this investigation, we revisit a controversial method of measuring transient extensional viscosity, originally pioneered by Doshi and Dealy in 1987; a method called “Exponential Shear”. From a purely continuum mechanics perspective, we develop an analytically tractable material function that quantitatively mimics planar extensional viscosity (derived analytically from a fully-nonlinear and frame-invariant continuum constitutive model), which appears uniformly valid over all Weissenberg numbers — and robust for a range of different viscoelastic fluids (including shear thinning, solvent viscosity, etc.). This material function requires only readily available experimental data from a bench-top, strain-controlled torsional rheometer. We demonstrate the method with data from a range of different complex fluids, and compare the results across a wide range of Weissenberg numbers. 

%We find that our analytic function unifies the debate from prior literature, accurately addressing the intermediate regime (Wi = 0.1 to 1). 

%Additionally, with modern Normal Stress sensors in the TA-instruments Ares-G2, we were able to experimentally reach finite extensibility in polymeric viscoelastic fluids — such an extreme limit that few extensional experimental techniques are capable of reaching. 

\end{quotation}

\onehalfspacing
\section{Introduction: A brief history of Large Amplitude Oscillatory Extension (LAOE) of complex fluids }

The cyclic \textit{extensional} properties of complex fluids are typically measured using custom experimental setups such as microfluidic methods \cite{zhou2016single, recktenwald2025large}. For soft solids, tensile mechanical testing units (Instron) or Dynamic Mechanical Analyzers (DMA) are used to obtain analogous measurements. Not only can such methods probe extensional properties in a single cycle (such as stiffness, viscous dissipation, or transient extensional viscosity), but critically, they can measure how these properties change over many cycles of successive loading. For soft solids, such techniques have become commonplace in industry; however, for fluids, the existing techniques are both challenging and have significant limitations. In this study, we propose a new approach for this type of fluid measurement, using only a commercial bench-top torsional rheometer. 

In 1987, \citet{doshi1987exponential} proposed a transient, exponential strain waveform in a homogeneous shear flow (i.e. cone and plate rheometer) of the form: 
\begin{equation}
    \gamma_{ES}(t) = 2 \sinh(\alpha t)
    \label{eq:ES}
\end{equation}

is capable of stretching viscoelastic material elements to locally mimic a steady planar extension rate, such that $\alpha = \dot{\varepsilon}$ (\citet{hencky1928} strain rate). It was recently shown in practice that this method can be used to accurately measure transient extensional viscosity across a diversity of fluid micro-structures (\citet{kroo2025unifying}). The key finding of this latest study was to compute the stretching rate (in the denominator of the extensional viscosity material function) to account for the \textit{affinity} of a material element (a quantity which can be well approximated with an estimated relaxation time). This study also identified a method to approximate this average relaxation time from the stress growth in a fluid -- by relating relaxation time to a characteristic timescale ($t^\star$) in exponential shear experimental data, when the first normal stress difference is  equal to twice the shear stress: $N_1(t^\star) = 2\sigma_{yx}(t^\star)$. This correction of the approximate stretching rate to account for affinity uses experimental data only (i.e. no model is used to estimate the relaxation time). 

Building off this necessary correction to exponential shear and the rigorous experimental validation of its practical implementation on a number of different complex fluids (including viscous fluids, quasi-linear viscoelastic fluids, nonlinear viscoelastic fluids, and shear-thinning fluids), we extend this technique to cyclic tensile measurements in fluids.  

\begin{figure*}
\label{fig:UCMresult}
\includegraphics[width=\textwidth]{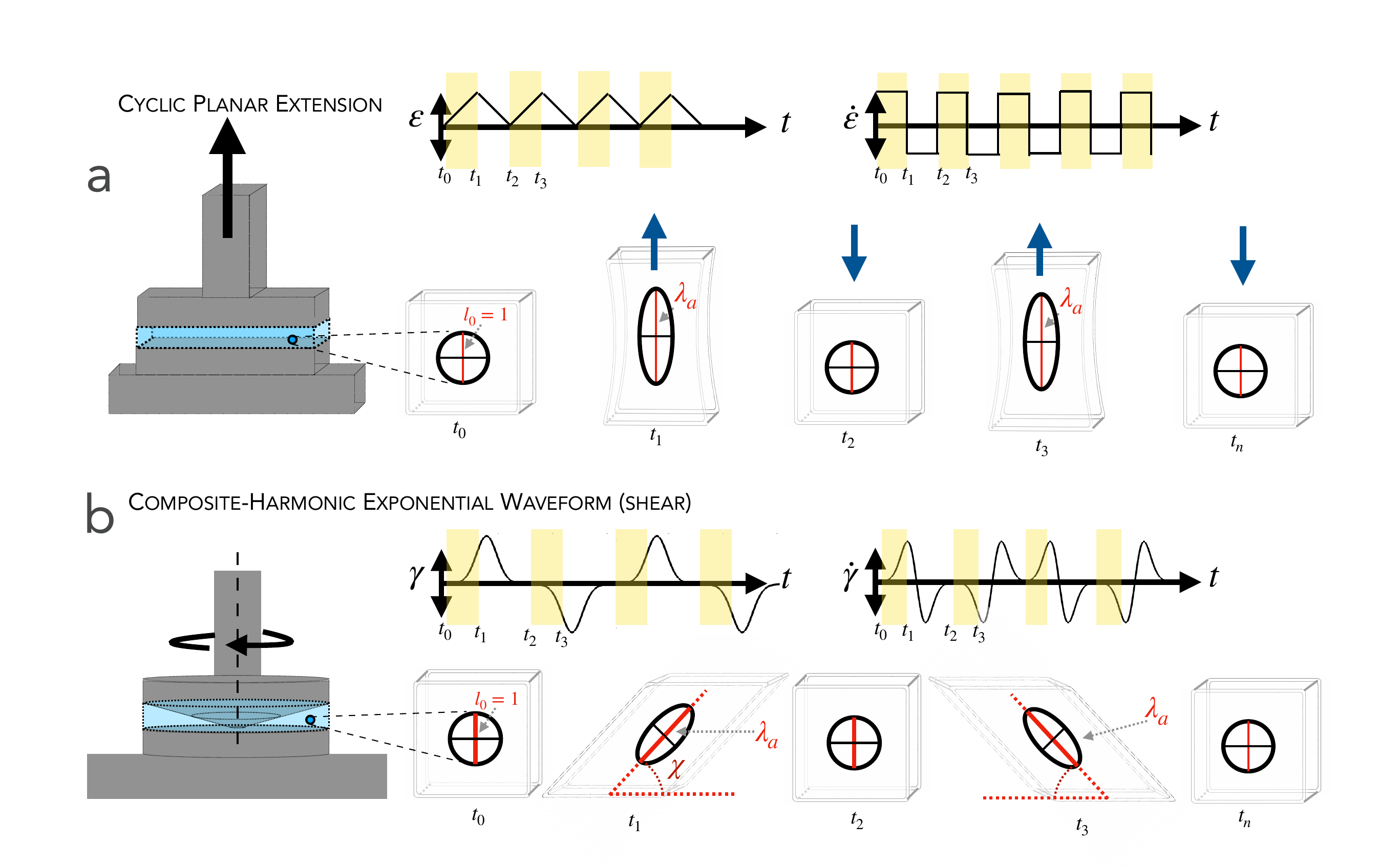}% Here is how to import EPS art
\caption{\label{fig:diagram} A new technique is presented that is designed to mimic a) A cyclic tensile stretching test in planar extension. (b) This new method uses a torsional rheometer to stretch elements of a homogeneous flow with a waveform that incorporates regions that are exponential in character -- while still having periodicity. The effect on a material element is shown in the second panel, where the material is cyclically stretched to mimic a constant Hencky strain rate over significant domains.}
\end{figure*}

The idea shown in figure \ref{fig:diagram}, is to use a torsional rheometer to mimic large amplitude cyclic planar extension of a fluid in a bulk sample under a homogeneous, prescribed time-varying strain. Each extension is steady such that $\dot{\varepsilon}$ is constant as shown in \ref{fig:diagram}a). This is accomplished by the construction of a new waveform input for a strain-controlled torsional rheometer(shown in figure\ref{fig:diagram}b): 

\begin{equation}
    \gamma_{CHEW}(t) = 2 \sinh(\hat{\alpha} \sin(\omega t))
    \label{eq:CHEW}
\end{equation}

This input is periodic but contains strongly exponential regions. Within specific windows (highlighted in yellow in the figure \ref{fig:diagram}b), the waveform is approximately equal to equation \ref{eq:ES}, and the transient extensional viscosity can be computed in these windows using the technique developed in \citet{kroo2025unifying}. In this manner, we can effectively accomplish measurement of rheometric properties of periodic stretching in a fluid without any specialized equipment; using only a commercial/bench-top strain-controlled shear rheometer. This fully-differentiable waveform allows the user to smoothly interpolate between exponential shear and small/medium/large amplitude oscillatory shear.  

\section{Kinematics }

\begin{figure*}[ht!]
\includegraphics[width=6.1 in]{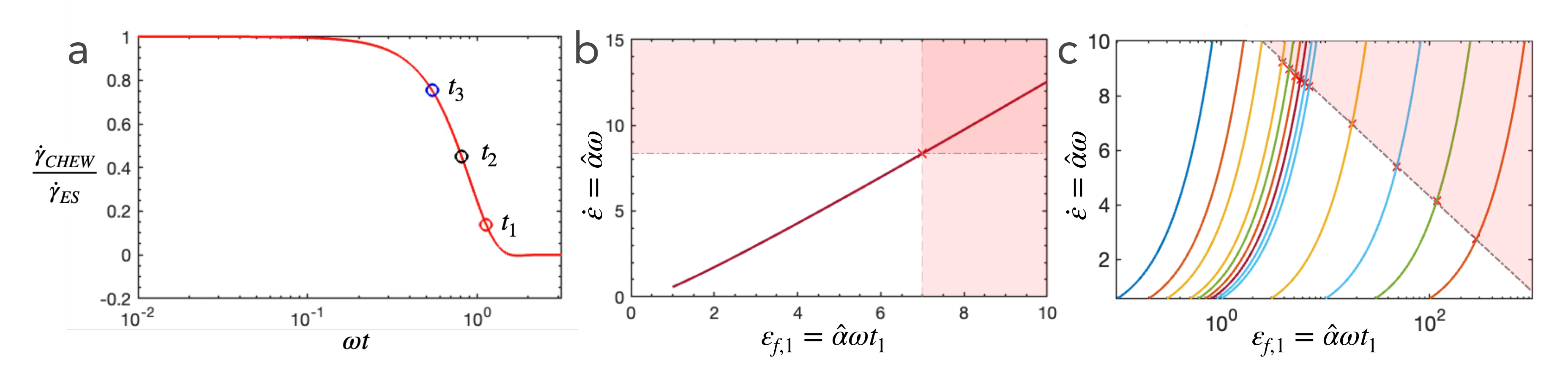}% Here is how to import EPS art
\caption{\label{fig:waveform} a)A comparison is shown between the cut-off criterion in equations 5-7. Here the ratio of the shear strain rate in CHEW versus true exponential shear is plotted. $t_1$ shows the least aggressive cut-off value. b) This waveform locks in a relationship between the effective Hencky strain \textit{amplitude} (evaluated at cutoff $t_1$), and the effective Hencky \textit{strain rate} for a given value of omega }
\end{figure*}

The strain input waveform described by equation \eqref{eq:CHEW} is described with 2 parameters: $\hat{\alpha}$, a dimensionless parameter that tunes both the flow-strength and flow-type simultaneously and $\omega$ (rad/s), the frequency. The waveform is shown in figure \ref{fig:waveform}a for different combinations of $\hat{\alpha}$ and $\omega$. 

The corresponding shear rate ($\dot{\gamma}$) is given by: 
\begin{equation}
    \dot{\gamma}_{CHEW} = 2\hat{\alpha}\cos{\omega t} \cosh{(\hat{\alpha} \sin{\omega t})}
    \label{eq:rate}
\end{equation}

This waveform is shown in figure \ref{fig:waveform}b for different values of $\hat{\alpha}$. 

For any value of $\hat{\alpha}$ or $\omega$, we can compute the  window where the waveform approximates true exponential shear (equation \eqref{eq:ES}). We define the non-dimensional cut-off time, $\omega t_n$, as the point at which the shear rate (or optionally, a higher order derivative) is at a local maxima. This cut-off point, where the CHEW waveform resembles exponential shear can be shown to be approximately: 
\begin{equation}
  \omega t_1 \approx \frac{\pi}{2} - \frac{1}{\sqrt{\hat{\alpha}}}
  \label{eq:cutoff}
\end{equation}

Prior to this cut-off time ($t_1$), the fluid is undergoing a classical exponential shear cycle, but where the effective Hencky strain rate ($\dot{\varepsilon}$) is given by $\hat{\alpha} \omega$ [s $^{-1}$] (rather than $\alpha$), and the effective final Hencky strain amplitude is given by $\varepsilon_f = \hat{\alpha} \omega t_1$ (rather than $\alpha t_1$). 

It is perhaps notable that the cutoff given in equation \ref{eq:cutoff} is a function of the parameter $\hat{\alpha}$. This implies that (for a given $\omega$ that sets the Hencky strain rate) the form of the CHEW function itself (eq \ref{eq:rate}) \textit{restricts} the maximum final strain amplitude. This is notable because the same is not strictly true of single-cycle exponential shear, where the duration of the experiment, $t_1$, is restricted only by the limitations of the motor, not by the choice of $\alpha$. Other extensional rheometric devices also do not typically have such a constraint. For example, on a filament stretching rheometer -- it is possible to arbitrarily set a constant Hencky strain rate independently from the achievable strain amplitude of the instrument. 

\begin{figure*}[ht!]
\includegraphics[width=\textwidth]{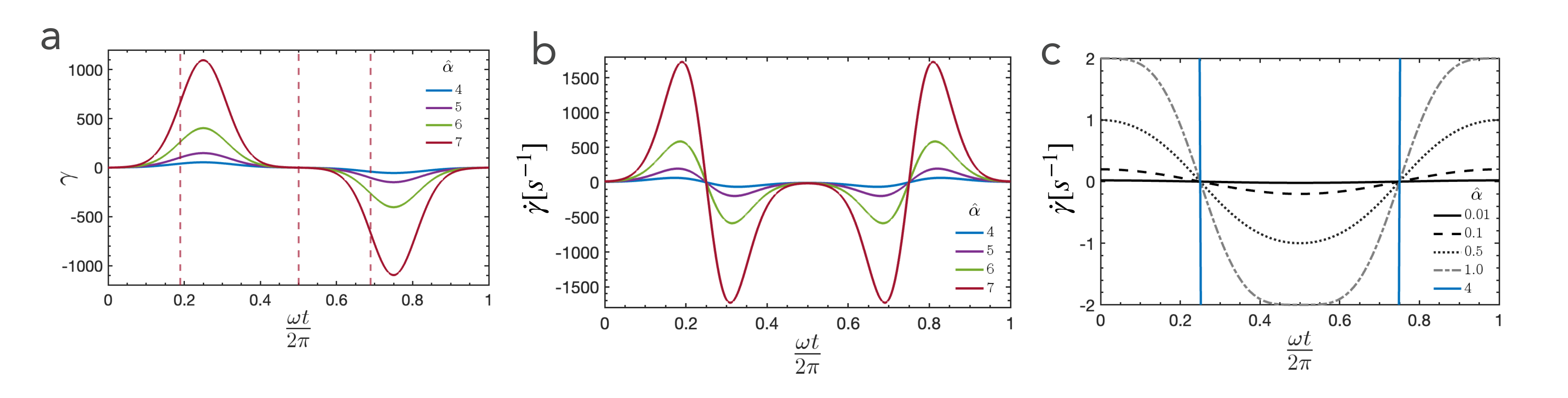}% Here is how to import EPS art
\caption{\label{fig:limitations}(a) The input strain is depicted for the composite harmonic exponential waveform (CHEW) described by eq.\ref{eq:CHEW}. The waveform is shown for $\omega = 1$ rad/s and $\hat{\alpha} = 1,2,3,4,5$. (b) The corresponding shear rate, described by eq. \ref{eq:rate} is depicted for the same values of $\omega$ and $\hat{\alpha}$
(c) The shear rate is depicted for small values of $\hat{\alpha} \omega$ ($0.01,0.1,0.5$, and $4 $s$^{-1})$, in the limit where the waveform approximates oscillatory shear rather than exponential shear.}

\end{figure*}

This can be visualized in Figure \ref{fig:limitations}c, where this relationship is visualized for $\omega = 1$ rad/s and $\hat{\alpha} = 1,2,3,4,5$. Further, we can also specify arbitrary and successively stricter criterion for  this cut-off time, such that the cut-off is defined by the local maxima of a derivative (of any order) of the strain (visualized in figure \ref{fig:limitations}a):

\begin{align}
t_1 &\coloneqq \left\{ t \,\middle|\, \frac{d^2 \gamma_{\text{CHEW}}}{dt^2} = 0 \right\} ~~(\mathrm{time~at~max~of~} \dot{\gamma}_{CHEW})\\
t_2 &\coloneqq \left\{ t \,\middle|\, \frac{d^3 \gamma_{\text{CHEW}}}{dt^3} = 0 \right\} ~(\mathrm{time~at~max~of~} \ddot{\gamma}_{CHEW})\\
t_3 &\coloneqq \left\{ t \,\middle|\, \frac{d^4 \gamma_{\text{CHEW}}}{dt^4} = 0 \right\}(\mathrm{time~at~max~of~} \dddot{\gamma}_{CHEW})
\end{align}

These criterion sets the relationship between the  achievable Hencky strain ($\varepsilon$) versus strain rate ($\dot{\varepsilon}$), as shown in Figure \ref{fig:limitations}b. It is notable that this effective Hencky strain rate ($\dot{\varepsilon} = \hat{\alpha} \omega$) sets the maximum shear rate in the material in an exponential manner, shown in Table \ref{table1}. Moderate Hencky strain rates quickly translate to  large cut-off transition shear rates at $t = t_1$. 

\begin{table}[h!]
\centering
\label{table1}
\begin{tabular}{c|c}
$\dot{\varepsilon}$ (s$^{-1}$) & $\left. \dot{\gamma} \right|_{t = t_1}$ (s$^{-1}$) \\
\hline
1 & 1.93 \\
2 & 6.227 \\
3 & 20.35 \\
4 & 64.22 \\
5 & 196.283 \\
6 & 586.9 \\
7 & 1728.4 \\
\end{tabular}
\caption{Relationship between $\dot{\varepsilon}$ and $\left. \dot{\gamma} \right|_{t = t_1}$}
\end{table}

\section{Experimental Examples}
\subsection{Newtonian Fluid}
\begin{figure*}

\includegraphics[width=\textwidth]{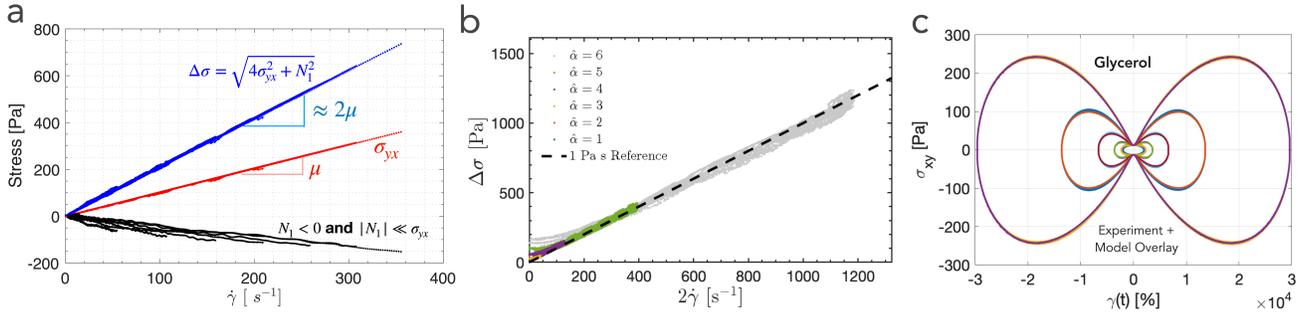}
\caption{\label{fig:newtonian} a)Stress components for a Newtonian oil b) Total principal stress versus two times the shear rate is shown for different values of $\hat{\alpha}$. c)Lissajous curves for this CHEW signal in stress versus strain take on a bilateral symmetry, due to the time-windows of exponential-like strain and the flow reversals inherent to the signal.}
\end{figure*}

This waveform when applied to a Newtonian calibration oil produces a small, negative, time-varying first normal stress difference ($N_1$) due to the inertia of the fluid, shown in Figure \ref{fig:newtonian}a. This small contribution from fluid inertia to $N_1$ is consistent with the results from \citet{kroo2025unifying} for exponential shear (and can be corrected  using the factor given by \citet{macosko1994rheology}
    $N_1 = N_{1,\mathrm{meas}} + 0.015 \cdot\rho\cdot\omega^2$).

A larger shear stress ($\sigma_{yx}$) also develops and is measured simultaneously (and independently). The total principal stress ($\Delta \sigma$ ) can be computed from these two independent, time-varying signals - shown for approximately 6 periods of CHEW at different values of $\hat{\alpha} \omega$ in figure \ref{fig:newtonian}b. 

This total principal stress is given by rotating the stress to the principal axes: 

\begin{equation}
    \Delta \sigma = \sqrt{4 \sigma_{yx}^2+N_1^2}
\end{equation}

We can show that (for $\omega$ = 1) as we vary the parameter $\hat{\alpha}$, the total principal stress divided by $2\dot{\gamma}$ results in a reasonably accurate description of the (rate-independent) viscosity for this Newtonian silicone oil. Viscosity fit values were obtained via linear regression and the standard deviation are computed for each effective Hencky strain rate, shown in table \ref{newtoniantable}) in Appendix \ref{Appendix A}. The standard deviation grows larger at larger $\hat{\alpha}$ values. 

\subsection{Second Order Fluid (PDMS)}
\begin{figure*}

\includegraphics[width=6.1 in]{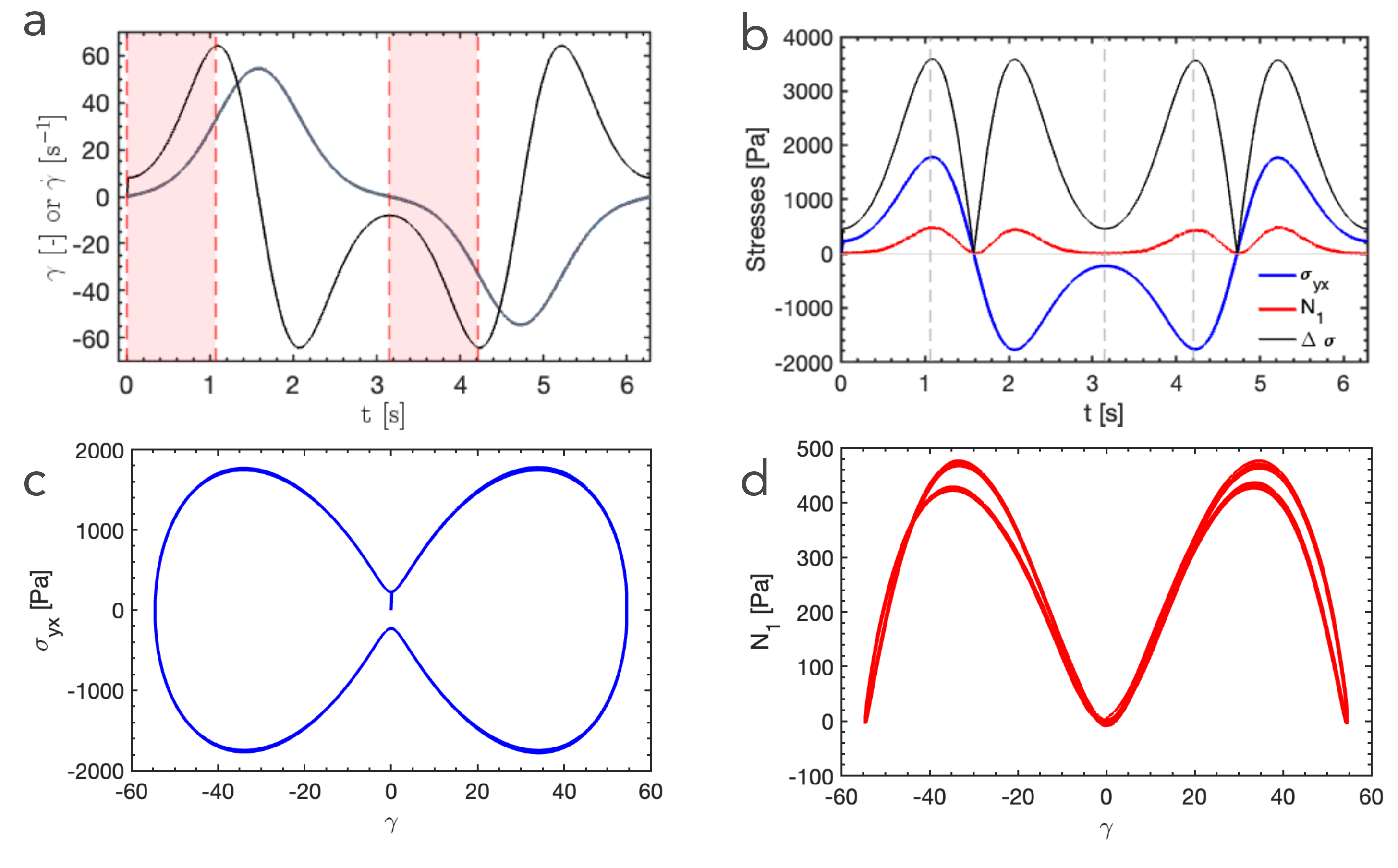}% Here is how to import EPS art
\caption{\label{fig:PDMS}a) PDMS demonstrates the example of a weakly elastic material in CHEW with $\hat{\alpha} = 4 \mathrm{~and~} \omega = 1$. b) The resulting stresses demonstrate a measurable first normal stress difference, and a much larger shear stress. Note that the maxima of the two stresses are approximately in-phase. c) The shear stress versus strain curve approximates this similar butterfly-like shape that is observable also in the Newtonian systems. d) The first normal stress difference versus strain has a collapsed form, enclosing very little area for this second-order fluid.   } 
\end{figure*}

For a second-order fluid, we demonstrate CHEW on a weakly viscoelastic polydimethylsiloxane (PDMS) sample (one of the same materials also characterized in \citet{kroo2025unifying} under single-cycle exponential shear). We observe the first normal stress difference is significantly smaller than the shear stress, and there appears to be no phase offset between the two stress components. The regions of exponential stretching of the material are highlighted in red in figure \ref{fig:PDMS}a. The total principal stress response to this cyclic stretching is depicted in figure \ref{fig:PDMS}b in black.  

\subsection{Nonlinear Elastic Fluid (PIB Boger Fluid)}
\begin{figure}

\includegraphics[width=\linewidth]{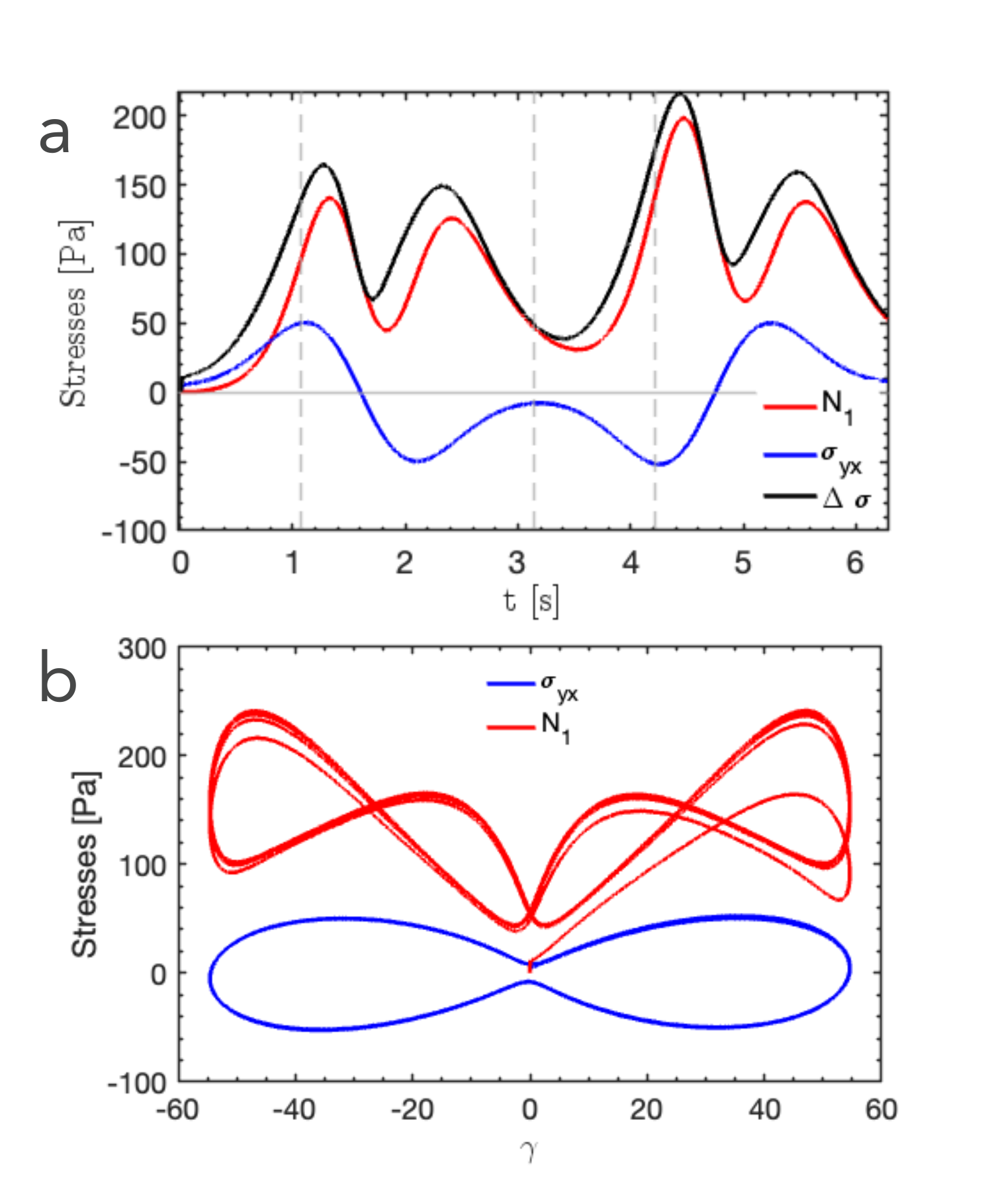}
\caption{\label{fig:PIB_BF}a)Stress components b) Lissajous curves for a PIB Boger fluid. $\dot{\varepsilon} = 4 $s$^{-1}$ and $\omega =1$ rad$/$sec.}
\end{figure}

For a PIB Boger fluid (in a polyalphaolefin / oil-based solvent), the stress components are shown -- as well as the total principal stress (in black) in response to a Hencky strain rate of $\varepsilon = 4$ s$^{-1}$. Because the longest relaxation time in this fluid is approximately $0.7$ s\cite{kroo2025unifying}, the Weissenberg number is $\approx 3$ -- classifying specific regions within transient flow as a "strong flow". In principle, this indicates that polymer molecules are likely stretching from a coiled to a stretched state during such flows. Note that the total stress is dominated by the first normal stress difference in this case, and a crossover is clearly shown in the stress components in figure \ref{fig:PIB_BF}a. %Polyalphaolefin (PAO) + PIB + oil base

\subsubsection{Transient Extensional Viscosity}
\begin{figure}
\includegraphics[width=\linewidth]{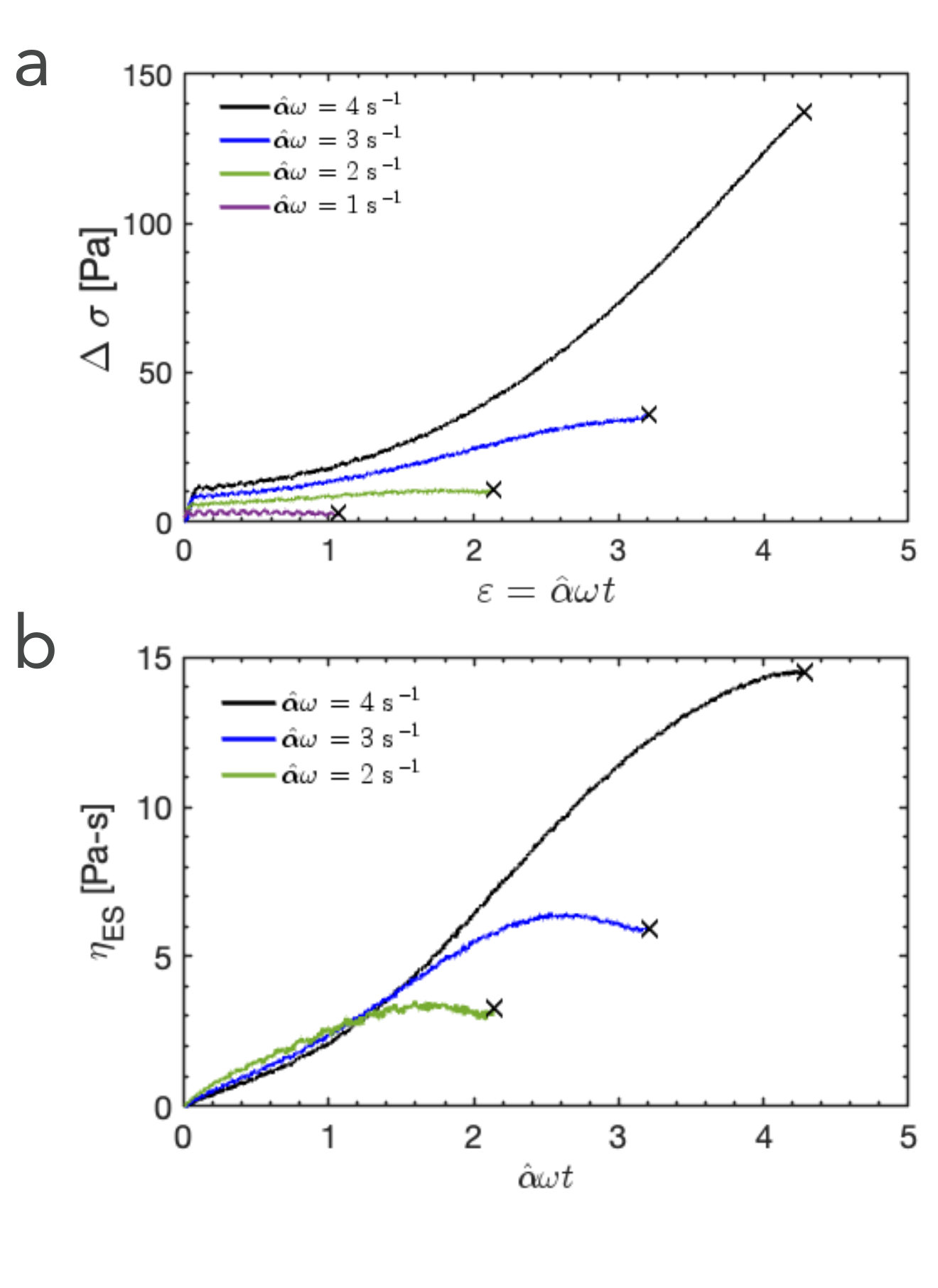}%
\caption{\label{fig:planarviscsoity} a) Total principal stress in the direction of stretch b) Transient planar extensional viscosity computed up until the cut-off value, $t_1$ according to the full updated method from \citet{kroo2025unifying}.} 

\end{figure}

Following the methods of \citet{kroo2025unifying}, we can compute the transient planar extensional viscosity using a Weissenberg-number dependent deformation rate to account for non-affine response of material elements (where the stress and strain tensors fail to be co-linear). We can apply this over a period of (approximate) stretching, choosing a cut-off criterion from equations 5-7. Here we choose $t_1$, and compute the extensional viscosity of this solution for Hencky strain rates up to $4$ s$^{-1}$. The principal stress is depicted in figure \ref{fig:planarviscsoity}a. As we increase the stretching rate, the extensional viscosity grows, indicative of dynamics of the coil-stretch transition and similar in many ways to the method single-cycle exponential shear\cite{kroo2025unifying}.

%similar to ES, there are transducer and motor limitations
% disimular to exponnetial shear, the t_f cutoff time is an appromximation. makes it look at those high amplitudes that the principle stress is rolling off. this is misleading. 
% potentially we need to use a more conservative cut-off time. heuristically, its about 67% of the cutoff. 

\section{Discussion: Limitations and Approximations}

The CHEW method, like exponential shear (ES)\cite{doshi1987exponential}, inherits several hardware-imposed limitations from the Ares-G2 (TA instruments) rheometer itself\cite{kroo2025unifying}. These include motor maximum shear rate limitations and transducer limits, all of which constrain the achievable Hencky strain rate and amplitude.

However, unlike ES, CHEW also imposes a built-in limitation due to the mathematical form of the waveform: a cutoff time $t_n$ used to define the extensional viscosity window is necessarily approximate. Because the waveform is cyclic, the strain amplitude is bounded by more than just the motor speed — and thus any definition of the "exponential-like" region requires a criterion for where exponential growth stops. In this study, we heuristically define the cutoff time as $t_1$, the point where $\dot{\gamma}_{\text{CHEW}}$ reaches a local maximum. However, in reality, there is a smooth transition away from the ideal waveform representing exponential shear, in all cases. 

This approximation can become misleading at high values of $\hat{\alpha}$, where the planar extensional viscosity appears to level-off near $t_1$ (due to this waveform curvature). A more conservative cutoff — defined using a second- or third-order maxima condition ($t_2$, $t_3$) — could help eliminate these effects but would significantly reduce the measurable strain amplitude. The main concern is that a user could interpret the measured stress output of the system as evidence of finite extensibility (when in fact, it is likely the waveform slowing down the effective material-element stretching rate as the signal approaches the cut-off criteria). 

Arguably, so long as this stress roll-off is not interpreted as evidence of finite extensibility -- the strong flow is still acting as a stretching flow on the fluid as the signal approaches $t_1$. An accurate description of this is perhaps that there is a departure from a true "steady planar extension rate". A better framing of this might be a stretching rate ($\varepsilon$) that starts off as effectively constant and then eventually falls off as a function of time. It is possible this (time-varying/subtly \textit{decelerating}) stretching flow may increase the residence time ($\varepsilon_f / \dot{\varepsilon}$) of polymer molecules at, say, a specific target conformation. This could be beneficial to some classes of processing flow.    

Lastly, similar to exponential shear as a technique, this method is not well suited for materials with very short relaxation times -- or fluids such as colloids known to produce negative normal stress differences. Likewise, we see that edge fracture may become problematic at high $\hat{\alpha} \omega$ values due to the aggressive flow reversal regions of the waveform. In these regions, the active stretching reverses and de-stretches (actively) -- not allowing the system to relax. While similar to cyclic tensile fatigue, the flow reversal itself may trigger tumbling events in polymeric molecules. This is a worthwhile future investigation, to understand the sensitivity of different polymeric fluids to flow reversal. 

We can also generalize the waveform in \ref{eq:CHEW} to scale the strain by a constant, A (rather than using 2) as shown in equation \ref{eq:newCHEW}):

\begin{equation}
    \gamma_{CHEW}(t) = A \sinh(\hat{\alpha} \sin(\omega t))
    \label{eq:newCHEW}
\end{equation}

While this technically departs from true planar extension, it still allows the user to quantitatively measure properties involved with material element stretching. Additionally, this allows the user to fix the initial strain rate of the waveform constant independently from $\hat{\alpha}$. 

\section{Evolving Materials}

\subsection{Damage-sensitive material: Provolone Cheese}
Fluid and soft-solid multiphase foodstuffs are well known to be highly susceptible to mechanical damage. Indeed, in food science, this property is often highly desirable, albeit challenging to measure. In the medical field of dysphagia (the condition associated with difficulty swallowing), measuring and controlling this property in soft-solid and liquid foods can be central to long-term patient outcomes. 

In the soft matter community, the effect of gradual and tuneable softening of a fluid (or soft-solid) microstructure upon successive cycling is typically called the "Mullins" effect, referring to the seminal studies of \citet{mullins1948effect} on the evolution of rubber elasticity to successive loading, in the 1940s.  

\begin{figure*}

\includegraphics[width=6.3 in]{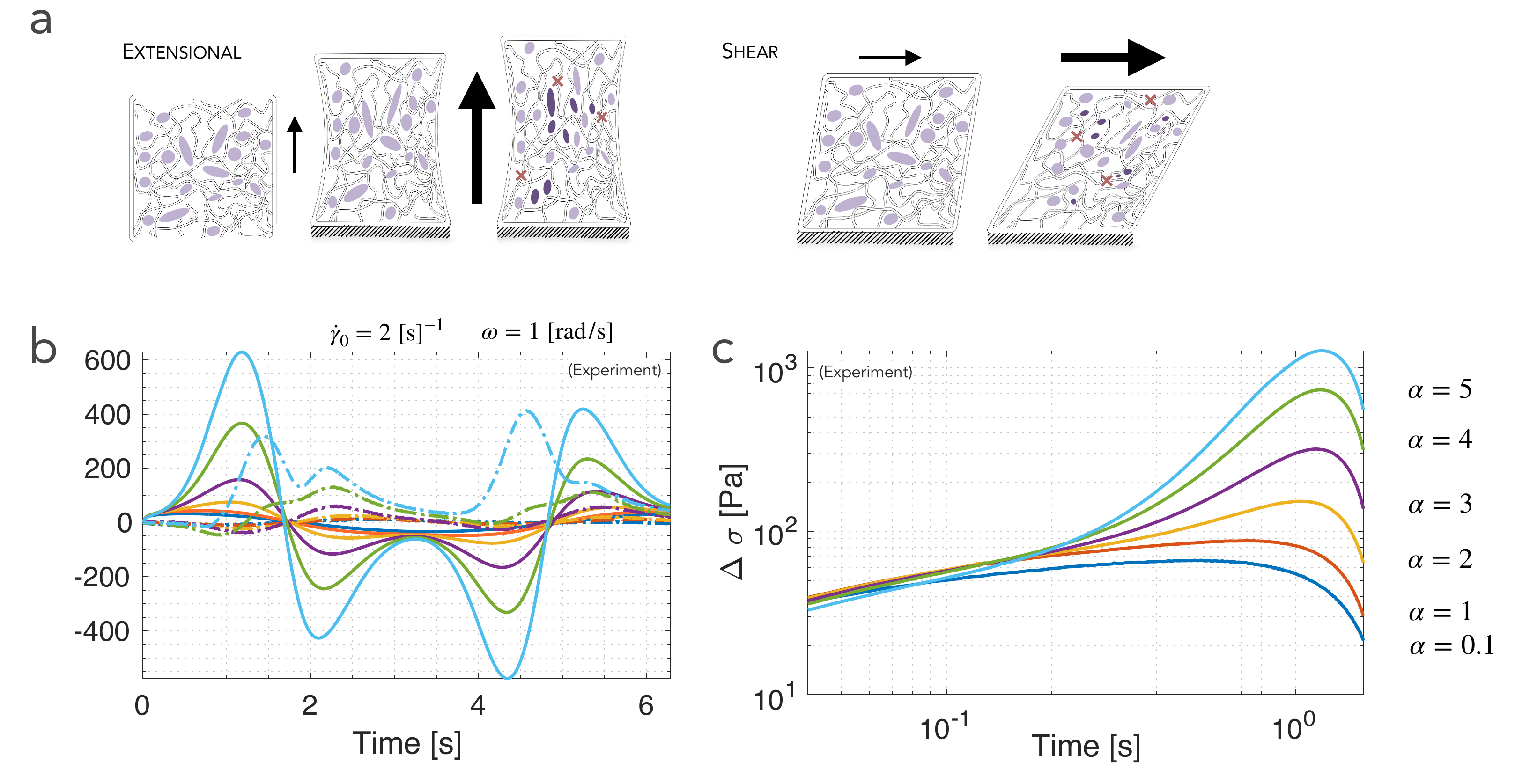}% Here is how to import EPS art
\caption{\label{fig:provolone1}(a)Conceptual depiction of the extension versus shear of a microstructure like melted cheese, containing significant complexity from multi-phase constituents and the transient protein network (b)Stress components for a melted provolone cheese sample at 65 degrees C. Solid lines represent shear stresses, dotted lines represent the normal stress difference. Values of $\hat{\alpha}$ are varied here from $0.1$ to $5$ s$^{-1}$(c) Total principal stress for the first quarter cycle of each CHEW experiment. }
\end{figure*}

\begin{figure*}

\includegraphics[width=6.3 in]{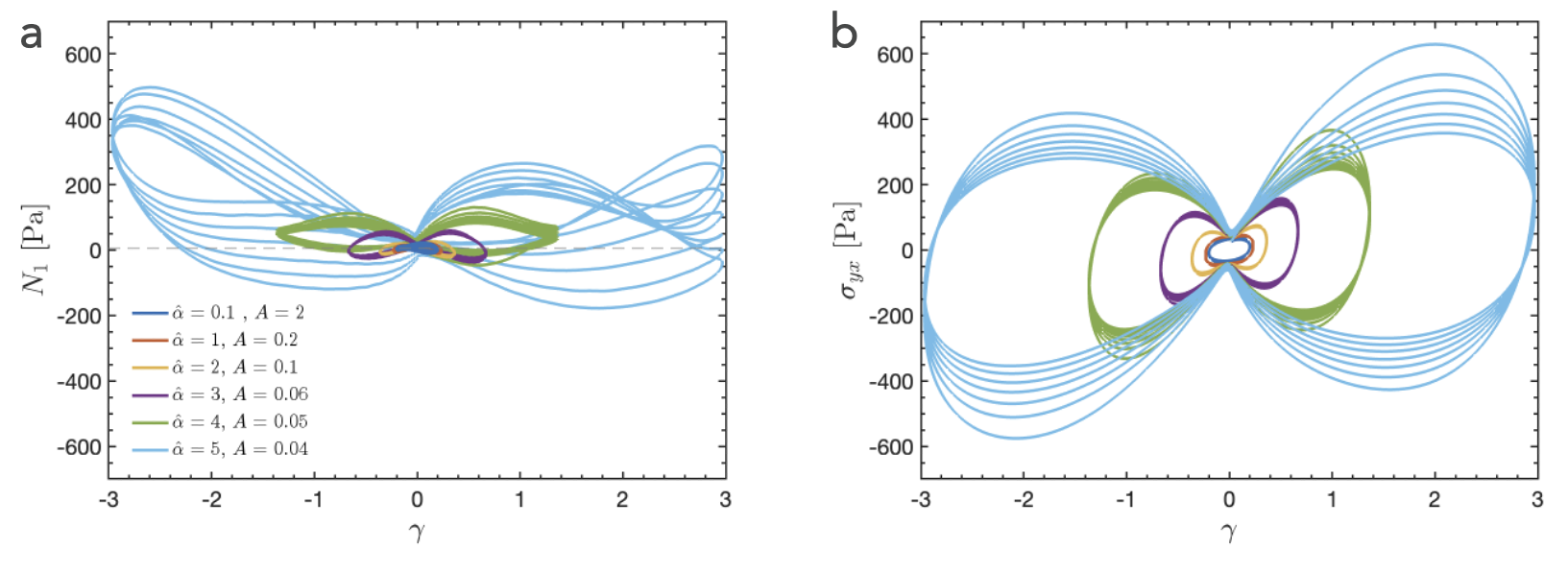}% Here is how to import EPS art
\caption{\label{fig:lissajousP} Lissajous curves for normal (a) versus shear (b) stress for a provolone cheese at 65 C. $\omega = 1$ rad/sec.}
\end{figure*}

\begin{figure*}
\label{fig:UCMresult}
\includegraphics[width=6.3 in]{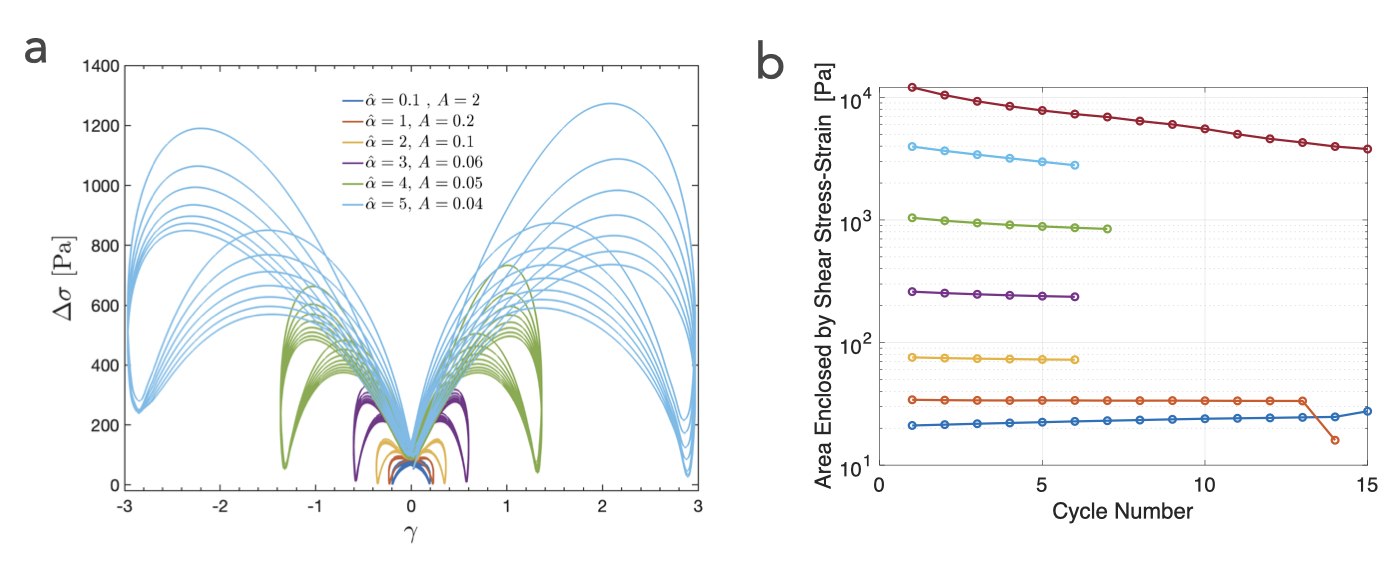}% Here is how to import EPS art
\caption{\label{fig:DS} (a) Lissajous curves for the total principal stress on provolone, exhibiting successive damage. (b) Quantifying this degradation, we can take metrics like area enclosed by these stress Lissajous curves to compute an energy dissipated per cycle.}
\end{figure*}

Here we show that the CHEW waveform can simultaneously measure and control the irreversible degradation of complex fluids. In figure \ref{fig:provolone1} we show the stress components and the total principal stress as a function of time for a number of different parameter regimes. This can also be interpreted as shown in figure \ref{fig:lissajousP}, showing the two lissajous curves (figure \ref{fig:lissajousP}a) of the first normal stress difference versus the shear stress (\ref{fig:lissajousP}b). These can be combined to compute a total principal stress (figure \ref{fig:DS}a). 

These trajectories in CHEW enclose areas on these lissajous curves, which can be characterized/analyzed as a quantification of the structural dissipation in each cycle. By computing the area and plotting the enclosed area over each successive cycle, it becomes clear that this waveform is capable of simultaneous precision sensing and actuation of structure. In the case of this foodstuff example, increasing effective Hencky strain rate tuneably degrades the material faster, cyle-over cycle (figure \ref{fig:DS}b).  

\subsection{Work-Hardening material: PVA-Borax}

\begin{figure*}

\includegraphics[width=6.3 in]{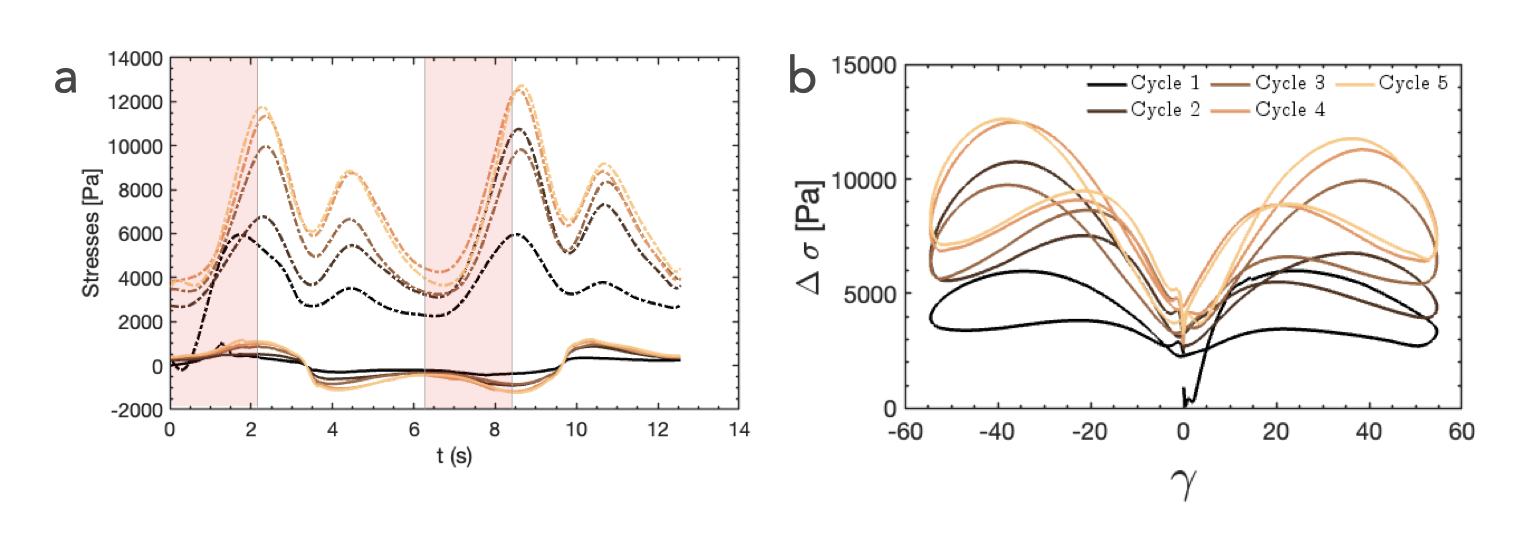}
\caption{\label{fig:PVA} a) Stress components for PVA borax b)Structuring of an associative polymer network with successive stretching cycles, shown with the Lissajous curves of total principal stress versus strain. }
\end{figure*}

Finally, we explore using CHEW to tunably structure materials upon cyclical stretching -- programming the assembly of fluids with sequences of stretching flow. Here we use a PVA-borax fluid, similar to \citet{ramlawi2024stress}, shown in figure \ref{fig:PVA}a. The regions in red are the regions of stretch. Each dotted line represents one successive cycle of normal stress, with the stresses growing significantly upon each cycle. We can also interpret this with the total principal stress (figure \ref{fig:PVA}b).   

%first quarter cycle is easy for the analysis -- just use the approximated tau from the method in the previous paper. , however, successive cycle dynamics are far harder because there is not a true crossing . How do you get the correct tau to use in the denominator of the material function? 

%interesting Q: do we really even need true eta_ES+? if not, just use principal stress. If yes, probably just use the estimated tau from the first cycle anyway. its likely sufficient.

% if you REALLY CARE about the precision-- the most obvious possibility -- one could just do sequences of single-cycle exponential shear to probe this. however, the role of relaxation makes the accumulated strain unclear. The benefit of CHEW is that there is kind of a constant/forced ACCUMULATED strain over each cycle -- like fatigue.  

% For successive cycles, one could use the phase offset between N1 and sigma_yx as a proxy for tau (similar to how the intersection was used in single cycle exponential shear)

% this is somewhat non-ideal as it only accounts for variations in tau between cycles. it is an approximation. It may be sufficient tho given that those offsets only change by XYZ% for the above examples, which only has a small effect on the actual extensional viscosity within a cycle (esp when plotted on loglog). 

\section{Discussion}

In this work, the analysis of the first quarter-cycle of the CHEW waveform is relatively straightforward, as it follows the method of \citet{kroo2025unifying} closely. Because this waveform approximates exponential shear within the first quarter-cycle stretching window, we can apply the method developed in Kroo et al. \cite{kroo2025unifying} to estimate the relaxation time $\tau$ and compute a transient extensional viscosity, $\eta^{+}_{ES}$, using a time-dependent principal stress.

However, interpretation of successive cycles literally as transient planar extensional viscosity is significantly more complicated. Unlike single-cycle exponential shear, CHEW has no clear or repeatable crossing where $N_1 = 2\sigma_{yx}$ after the first cycle. As such, there is no obvious, cycle-specific way to define the relaxation time $\tau$, which appears in the denominator of the extensional viscosity material function.

This raises a broader question: is a precise estimate of $\eta^{+}_{ES}$ in subsequent cycles truly necessary? In many cases, the principal stress trajectory itself may be sufficiently descriptive, especially if the user is primarily interested in the \textit{evolution} of material response -- or in relative changes across cycles.

However, for applications where high accuracy is required (and $\tau$ changes significantly, as we suspect is the case with our mutating material examples), we must explore are alternative strategies. 

A promising direction to estimate $\tau$ in successive cycles is to use the phase offset between the first normal stress difference ($N_1(t)$) and the shear stress ($\sigma_{yx}(t)$). In principle, this offset encodes information about viscoelastic lag and can serve as a proxy for $\tau$, just as the crossing between $N_1$ and $\sigma_{yx}$ was used in exponential shear. This approach represents one possible promising direction for more precise quantitative interpretations of CHEW waveforms in future.

In summary, while CHEW introduces analytical challenges beyond the first cycle, the technique retains several practical advantages. It allows for a natural accumulation of strain, akin to cyclic fatigue in solids, and it produces interpretable stress signals -- even the practical ability to define an approximate extensional viscosity (especially for fluids that do not mutate dramatically). The balance between approximation and utility will likely depend on the specific application.

\section*{References}

\nocite{*}
\bibliography{aipsamp}% Produces the bibliography via BibTeX.

\begin{thebibliography}{8}%
\makeatletter
\providecommand \@ifxundefined [1]{%
 \@ifx{#1\undefined}
}%
\providecommand \@ifnum [1]{%
 \ifnum #1\expandafter \@firstoftwo
 \else \expandafter \@secondoftwo
 \fi
}%
\providecommand \@ifx [1]{%
 \ifx #1\expandafter \@firstoftwo
 \else \expandafter \@secondoftwo
 \fi
}%
\providecommand \natexlab [1]{#1}%
\providecommand \enquote  [1]{``#1''}%
\providecommand \bibnamefont  [1]{#1}%
\providecommand \bibfnamefont [1]{#1}%
\providecommand \citenamefont [1]{#1}%
\providecommand \href@noop [0]{\@secondoftwo}%
\providecommand \href [0]{\begingroup \@sanitize@url \@href}%
\providecommand \@href[1]{\@@startlink{#1}\@@href}%
\providecommand \@@href[1]{\endgroup#1\@@endlink}%
\providecommand \@sanitize@url [0]{\catcode `\\12\catcode `\$12\catcode `\&12\catcode `\#12\catcode `\^12\catcode `\_12\catcode `\%12\relax}%
\providecommand \@@startlink[1]{}%
\providecommand \@@endlink[0]{}%
\providecommand \url  [0]{\begingroup\@sanitize@url \@url }%
\providecommand \@url [1]{\endgroup\@href {#1}{\urlprefix }}%
\providecommand \urlprefix  [0]{URL }%
\providecommand \Eprint [0]{\href }%
\providecommand \doibase [0]{http://dx.doi.org/}%
\providecommand \selectlanguage [0]{\@gobble}%
\providecommand \bibinfo  [0]{\@secondoftwo}%
\providecommand \bibfield  [0]{\@secondoftwo}%
\providecommand \translation [1]{[#1]}%
\providecommand \BibitemOpen [0]{}%
\providecommand \bibitemStop [0]{}%
\providecommand \bibitemNoStop [0]{.\EOS\space}%
\providecommand \EOS [0]{\spacefactor3000\relax}%
\providecommand \BibitemShut  [1]{\csname bibitem#1\endcsname}%
\let\auto@bib@innerbib\@empty
%</preamble>
\bibitem [{\citenamefont {Kroo}\ \emph {et~al.}(2025)\citenamefont {Kroo}, \citenamefont {Nicholson}, \citenamefont {Boehm}, \citenamefont {Baier},\ and\ \citenamefont {McKinley}}]{kroo2025unifying}%
  \BibitemOpen
  \bibfield  {author} {\bibinfo {author} {\bibfnamefont {L.~A.}\ \bibnamefont {Kroo}}, \bibinfo {author} {\bibfnamefont {R.~A.}\ \bibnamefont {Nicholson}}, \bibinfo {author} {\bibfnamefont {M.~W.}\ \bibnamefont {Boehm}}, \bibinfo {author} {\bibfnamefont {S.~K.}\ \bibnamefont {Baier}}, \ and\ \bibinfo {author} {\bibfnamefont {G.~H.}\ \bibnamefont {McKinley}},\ }\href@noop {} {\enquote {\bibinfo {title} {A unifying perspective on measuring transient planar extensional viscosity from exponential shear},}\ } (\bibinfo {year} {2025}),\ \bibinfo {note} {\href{http://laurel.kroo.net/s/Kroo_2025a}{ arXiv Preprint \texttt{arXiv:submit/6556503} [cond-mat.soft]}}\BibitemShut {NoStop}%
\bibitem [{\citenamefont {Zhou}\ and\ \citenamefont {Schroeder}(2016)}]{zhou2016single}%
  \BibitemOpen
  \bibfield  {author} {\bibinfo {author} {\bibfnamefont {Y.}~\bibnamefont {Zhou}}\ and\ \bibinfo {author} {\bibfnamefont {C.~M.}\ \bibnamefont {Schroeder}},\ }\bibfield  {title} {\enquote {\bibinfo {title} {Single polymer dynamics under large amplitude oscillatory extension},}\ }\href@noop {} {\bibfield  {journal} {\bibinfo  {journal} {Physical Review Fluids}\ }\textbf {\bibinfo {volume} {1}},\ \bibinfo {pages} {053301} (\bibinfo {year} {2016})}\BibitemShut {NoStop}%
\bibitem [{\citenamefont {Recktenwald}\ \emph {et~al.}(2025)\citenamefont {Recktenwald}, \citenamefont {John}, \citenamefont {Shen}, \citenamefont {Poole}, \citenamefont {Fonte},\ and\ \citenamefont {Haward}}]{recktenwald2025large}%
  \BibitemOpen
  \bibfield  {author} {\bibinfo {author} {\bibfnamefont {S.~M.}\ \bibnamefont {Recktenwald}}, \bibinfo {author} {\bibfnamefont {T.~P.}\ \bibnamefont {John}}, \bibinfo {author} {\bibfnamefont {A.~Q.}\ \bibnamefont {Shen}}, \bibinfo {author} {\bibfnamefont {R.~J.}\ \bibnamefont {Poole}}, \bibinfo {author} {\bibfnamefont {C.~P.}\ \bibnamefont {Fonte}}, \ and\ \bibinfo {author} {\bibfnamefont {S.~J.}\ \bibnamefont {Haward}},\ }\bibfield  {title} {\enquote {\bibinfo {title} {Large amplitude oscillatory extension (laoe) of dilute polymer solutions},}\ }\href@noop {} {\bibfield  {journal} {\bibinfo  {journal} {arXiv preprint arXiv:2501.11950}\ } (\bibinfo {year} {2025})}\BibitemShut {NoStop}%
\bibitem [{\citenamefont {Doshi}\ and\ \citenamefont {Dealy}(1987)}]{doshi1987exponential}%
  \BibitemOpen
  \bibfield  {author} {\bibinfo {author} {\bibfnamefont {S.}~\bibnamefont {Doshi}}\ and\ \bibinfo {author} {\bibfnamefont {J.}~\bibnamefont {Dealy}},\ }\bibfield  {title} {\enquote {\bibinfo {title} {Exponential shear: a strong flow},}\ }\href@noop {} {\bibfield  {journal} {\bibinfo  {journal} {Journal of Rheology}\ }\textbf {\bibinfo {volume} {31}},\ \bibinfo {pages} {563--582} (\bibinfo {year} {1987})}\BibitemShut {NoStop}%
\bibitem [{\citenamefont {Hencky}(1928)}]{hencky1928}%
  \BibitemOpen
  \bibfield  {author} {\bibinfo {author} {\bibfnamefont {H.}~\bibnamefont {Hencky}},\ }\bibfield  {title} {\enquote {\bibinfo {title} {{\"U}ber die form des elastizit{\"a}tsgesetzes bei ideal elastischen stoffen},}\ }\href@noop {} {\bibfield  {journal} {\bibinfo  {journal} {Zeitschrift f{\"u}r technische Physik}\ }\textbf {\bibinfo {volume} {9}},\ \bibinfo {pages} {215--220} (\bibinfo {year} {1928})}\BibitemShut {NoStop}%
\bibitem [{\citenamefont {Macosko}(1994)}]{macosko1994rheology}%
  \BibitemOpen
  \bibfield  {author} {\bibinfo {author} {\bibfnamefont {C.~W.}\ \bibnamefont {Macosko}},\ }\href {https://www.amazon.com/Rheology-Measurements-Applications-Christopher-Macosko/dp/0471185752} {\emph {\bibinfo {title} {Rheology: Principles, Measurements, and Applications}}}\ (\bibinfo  {publisher} {Wiley},\ \bibinfo {year} {1994})\BibitemShut {NoStop}%
\bibitem [{\citenamefont {Mullins}(1948)}]{mullins1948effect}%
  \BibitemOpen
  \bibfield  {author} {\bibinfo {author} {\bibfnamefont {L.}~\bibnamefont {Mullins}},\ }\bibfield  {title} {\enquote {\bibinfo {title} {Effect of stretching on the properties of rubber},}\ }\href@noop {} {\bibfield  {journal} {\bibinfo  {journal} {Rubber chemistry and technology}\ }\textbf {\bibinfo {volume} {21}},\ \bibinfo {pages} {281--300} (\bibinfo {year} {1948})}\BibitemShut {NoStop}%
\bibitem [{\citenamefont {Ramlawi}\ \emph {et~al.}(2024)\citenamefont {Ramlawi}, \citenamefont {Hossain}, \citenamefont {Shetty},\ and\ \citenamefont {Ewoldt}}]{ramlawi2024stress}%
  \BibitemOpen
  \bibfield  {author} {\bibinfo {author} {\bibfnamefont {N.}~\bibnamefont {Ramlawi}}, \bibinfo {author} {\bibfnamefont {M.~T.}\ \bibnamefont {Hossain}}, \bibinfo {author} {\bibfnamefont {A.}~\bibnamefont {Shetty}}, \ and\ \bibinfo {author} {\bibfnamefont {R.~H.}\ \bibnamefont {Ewoldt}},\ }\bibfield  {title} {\enquote {\bibinfo {title} {Stress-controlled medium-amplitude oscillatory shear (maostress) of pva--borax},}\ }\href@noop {} {\bibfield  {journal} {\bibinfo  {journal} {Journal of Rheology}\ }\textbf {\bibinfo {volume} {68}},\ \bibinfo {pages} {741--763} (\bibinfo {year} {2024})}\BibitemShut {NoStop}%
\end{thebibliography}%
\section{Appendix A: Newtonian Data}
Here we show the linear regression analysis for the Newtonian data in figure \ref{fig:newtonian}b. Note that for 6 cycles of CHEW, the average viscosity at all tested values is quite close to the 1 Pa-s standard used. 
\label{Appendix A}
\begin{table}[h!]
\centering
\begin{tabular}{c c c c}
\hline
$\hat{\alpha}$ & Visc fit (Pa$\cdot$s) & STD (Residual) & \% Error \\
\hline
1 & 1.0782 & 1.4075   & 7.8\% \\
2 & 1.0586 & 1.3012   & 5.9\% \\
3 & 1.1143 & 4.6131   & 11.4\% \\
4 & 1.1525 & 13.1180  & 15.3\% \\
5 & 1.0375 & 24.3496  & 3.7\% \\
6 & 0.9866 & 29.4097  & 1.4\% \\
\hline
\end{tabular}
\caption{\label{newtoniantable}Viscosity fit results for different values of $\hat{\alpha}$ for a viscosity standard silicone oil.}
\end{table}
\end{document}